\documentclass[superscriptaddress,twocolumn]{revtex4-1}

\usepackage{graphicx}
\usepackage{amsmath}
\usepackage{longtable}
\usepackage{amssymb}
\usepackage{latexsym}
\usepackage{natbib}

\usepackage{times}
\usepackage{graphics}
\usepackage{graphicx}
\usepackage{epsfig}
\usepackage{wrapfig}
\usepackage[T1]{fontenc}
\usepackage{float}
\usepackage{enumerate}
\usepackage{setspace,ifthen}
\usepackage{amsthm} 
\usepackage{amssymb}	
\usepackage[usenames,dvipsnames]{xcolor}
\usepackage{mathrsfs,amsmath}
\usepackage{mathrsfs}
\usepackage{mathrsfs}
\usepackage{rotating}
\usepackage{epstopdf}
\usepackage{appendix}
\usepackage{kbordermatrix} 
\usepackage{blkarray}
\usepackage{multirow}

\usepackage{array}

\usepackage{rotate}

\usepackage{kbordermatrix} 

\renewcommand{\kbldelim}{[} 
\renewcommand{\kbrdelim}{]}
\usepackage{braket}

\newcommand{\abs}[1]{\left| #1 \right|} 
 
\let\baraccent=\= 
\renewcommand{\=}[1]{\stackrel{#1}{=}} 

\theoremstyle{definition}

\theoremstyle{remark}

\newcommand{\blu}{\color{blue}}

\newcommand{\sigz}{\hat{\sigma}_z}

\newcommand{\sigx}{\hat{\sigma}_x}
\newcommand{\sigy}{\hat{\sigma}_y}

\newcommand{\BinaryTS}{\sigma}
\newcommand{\BinaryVecTS}{\vec{\boldsymbol{\BinaryTS}}}

\RequirePackage[normalem]{ulem} 
\RequirePackage{color}\definecolor{RED}{rgb}{1,0,0}\definecolor{BLUE}{rgb}{0,0,1} 

\begin{document}



\title{A functional basis for efficient physical-layer classical control in quantum processors}

\author{Harrison Ball}
\affiliation{ARC Centre of Excellence for Engineered Quantum Systems, School of Physics, The University of Sydney, NSW 2006 Australia}
\author{Trung Nguyen}
\affiliation{School of Electrical and Information Engineering, The University of Sydney, NSW 2006 Australia}
\affiliation{ARC Centre of Excellence for Engineered Quantum Systems, School of Physics, The University of Sydney, NSW 2006 Australia}
\author{Philip H. W. Leong}
\affiliation{School of Electrical and Information Engineering, The University of Sydney, NSW 2006 Australia}
\author{Michael J. Biercuk}
\email[To whom correspondence should be addressed: ]{michael.biercuk@sydney.edu.au}
\affiliation{ARC Centre of Excellence for Engineered Quantum Systems, School of Physics, The University of Sydney, NSW 2006 Australia}

\date{\today}



\begin{abstract}
The rapid progress seen in the development of quantum coherent devices for information processing has motivated serious consideration of quantum computer architecture and organization.  One topic which remains open for investigation and optimization relates to the design of the classical-quantum interface, where control operations on individual qubits are applied according to higher-level algorithms; accommodating competing demands on performance and scalability remains a major outstanding challenge.  In this work we present a resource-efficient, scalable framework for the implementation of embedded physical-layer classical controllers for quantum information systems.  Design drivers and key functionalities are introduced, leading to the selection of Walsh functions as an effective functional basis for both programming and controller hardware implementation. This approach leverages the simplicity of real-time Walsh-function generation in classical digital hardware, and the fact that a wide variety of physical-layer controls such as dynamic error suppression are known to fall within the Walsh family.  We experimentally implement a real-time FPGA-based Walsh controller producing Walsh timing signals and Walsh-synthesized analog waveforms appropriate for critical tasks in error-resistant quantum control and noise characterization.  These demonstrations represent the first step towards a unified framework for the realization of physical-layer controls compatible with large-scale quantum information processing. 

\end{abstract}


\maketitle


\section{Introduction}
The suppression of error in quantum systems is a key consideration in many studies of quantum computer architecture~\cite{meter2006architectural, van2008architecture, DiVincenzoArch, Fowler2010, JonesPRX2012, Fowler_Surface}, and emerged early in the community~\cite{oskin2002practical} as a significant driver for system design.  Unsurprisingly, many studies have focused on the extreme demands in quantum device performance and allocation imparted by logical encoding in quantum error correction (QEC), revealing substantial resource demands on the path to fault-tolerance.  Simultaneously, open-loop dynamic error suppression (DES) techniques~\cite{Viola1998,Viola1999,Zanardi1999,Vitali1999,Viola2003,Byrd2003,Kofman2004,Khodjasteh2005,Yao2007,Uhrig2007,Gordon2008,Khodjasteh2009dcg,Khodjasteh2009,Khodjasteh2010,Liu2010,Biercuk_Filter} operating at the physical layer have widely been identified as an important potential complement to QEC~\cite{Preskill_Layered,khodjasteh_fault-tolerant_2005, khodjasteh_rigorous_2008, JonesPRX2012}.  This is in part because of their potential to improve the resource-efficiency of QEC encoding by reducing the physical-qubit error rate.  We expect that in general, DES-operations will be implemented at the physical level for effectively all physical qubit manipulation.

DES techniques have now been experimentally demonstrated to be highly effective~\cite{BiercukNature2009, Du2009, Bluhm, Gustavsson2012, SoareNatPhys2014, Marcus2016}, but are themselves complex, and consideration of overhead and resource allocation associated with their application is also necessary.  For instance, general DES protocols require  classical control hardware appropriate for the generation of precisely timed and temporally modulated sequences of physical operations tailored to the underlying hardware noise model.  The performance specifications of these classical controllers (e.g. in timing resolution or latency~\cite{Hodgson2010, Biercuk_Filter}) become increasingly demanding as net error suppression requirements increase.  

The bulk of today's control-hardware solutions employing high-performance but high-cost waveform generators at the benchtop are not scalable.  Meanwhile, custom electronics solutions~\cite{lamb2016fpga, hornibrook2015} have focused on reproducing conventional control capabilities in streamlined hardware.  To date, studies in this space have not addressed the key opportunity to efficiently allocate classical control hardware resources by considering how to leverage the underlying physics of the control operations themselves.  

In this manuscript we address this challenge by first proposing a functional basis for physical-layer classical controllers compatible with both conventional digital electronics and a wide variety of quantum control protocols.  Our analysis identifies the Walsh functions~\cite{Walsh_Tzafestas, Walsh_Beauchamp} as providing a critical mathematical link between the physics of quantum control and the requisite physical implementation of control timing, gate implementation, modulation-waveform synthesis, and environmental noise spectroscopy.  We build on a growing body of literature focusing on Walsh functions in quantum control~\cite{Leung:1999, Leung2002, HayesPRA2011, HayesPRL, Ball2014, SoareNatPhys2014, Cooper14, Green2014, AAGuzik_2014, LongStorage}, and consider here how the use of Walsh functions as a functional programming basis dramatically simplifies the design of embedded classical controllers at the physical-layer.  In order to demonstrate the benefits of working within the overarching Walsh framework, we build customized real-time electronic modules in a field-programmable gate array (FPGA) and demonstrate critical functionality including: Walsh-function generation, sequencing based on Walsh-timing information, Walsh-synthesis of analog DES modulation waveforms, and Walsh-based noise spectroscopy and signal reconstruction.  These functions demonstrate the core capabilities of a high-efficiency classical controller at the physical layer, and are shown to outperform generic signal synthesis in an embedded microprocessor by factors of $>10^{5}$ in terms of signal latency and required computational resources.  Our demonstrations highlight a path to realize scalable, ``lightweight,'' embedded control solutions providing high-performance quantum control capabilities.  

The remainder of this manuscript is organized as follows. We begin by introducing the relevance of Walsh functions for physical-layer control of quantum information systems in Sec.~\ref{Sec:Walsh}.  We then move on to articulate key \emph{Summary Requirements} for physical-layer controller functionality in the Walsh framework as well as \emph{Design Drivers} constraining potential solutions in Sec.~\ref{Sec:FPGA}, followed by an experimental implementation of a resource-efficient real-time embedded controller addressing the desiderata laid out earlier in the manuscript.  We present the relevant controller architecture, demonstrations of representative output, and a comparative analysis relative to a simple microcontroller-based implementation before providing a summary and conclusion in Sec.~\ref{Sec:Conclusion}.

\section{Walsh functions a functional programming basis for physical-layer controllers}\label{Sec:Walsh}

Physical layer controllers have the role of implementing control operations in a manner invisible to higher level operations, effectively constituting a type of quantum firmware.  This firmware is designed to enact manipulations on physical qubits according to sequencing determined in an abstracted algorithm, and in a manner compatible with the scaling demands of quantum error correction.  This implies that physical-layer controls must be robust against ambient and control-based noise sources, and even capable of suppressing error below fault-tolerance thresholds.  Our objective here is to understand how these functional requirements, combined with system-level constraints, drive the design of classical controllers at the physical level, and how such design considerations in turn affect the choice of control framework.

Contemporary system-level analyses of quantum computer architecture~\cite{JonesPRX2012} make clear that there will be a need for dedicated classical control hardware, often discussed in a physically distributed hierarchy~\cite{lamb2016fpga, hornibrook2015}, with dedicated local resources at the physical-qubit layer.  In this generalized framework electronics operating at the physical level can be thought of in the context of \emph{embedded} controllers, bringing with them a new class of specific design constraints.  Such embedded controllers must operate semi-autonomously within a hierarchical control architecture, applying and managing complex physical-layer protocols on demand from a master controller operating at higher levels of system abstraction.

In the most naive approach, one may consider implementing physical-layer controllers using standard microcontrollers generating and outputting relevant control signals.  However, achieving broad flexibility of the controller's functionality through such general-purpose hardware leads to degraded performance;  the controller effectively provides unnecessary capability to the system (e.g. not all possible signal output patterns will be necessary to control qubits), and simultaneously introduces shortcomings in terms of latency and the complexity of programming the device relative to application-specific designs.    

We ask the question whether it is possible to constrain the space of available controls in such a way as to both provide all necessary physical-layer control and sequencing functionality and simplify the underlying classical hardware.  Our analysis answers this question in the affirmative; the identification of an appropriate mathematical framework can provide a critical link between a (truncated) space of allowed quantum control protocols and sequences and the architecture of the associated hardware. Here, we identify the Walsh functions as a simple encoding basis for the the physical layer controller, bridging the hardware-software interface for quantum control.  This family of bi-valued piecewise-constant functions, Fig.~\ref{Fig:Walsh}a, has seen a large variety of applications in digital signal processing~\cite{Walsh_Beauchamp,Walsh_Tzafestas}, and has more recently been adopted in the fields of quantum control and quantum information (see Appendix~\ref{Ap:WalshOverview} for a full summary). In our analyses we consider the Walsh functions as a single unifying mathematical and programming basis for physical-layer control in quantum information processors, leveraging the following key strengths: 

\begin{figure}[tp]
      \centering
      \includegraphics[width=0.9\columnwidth]{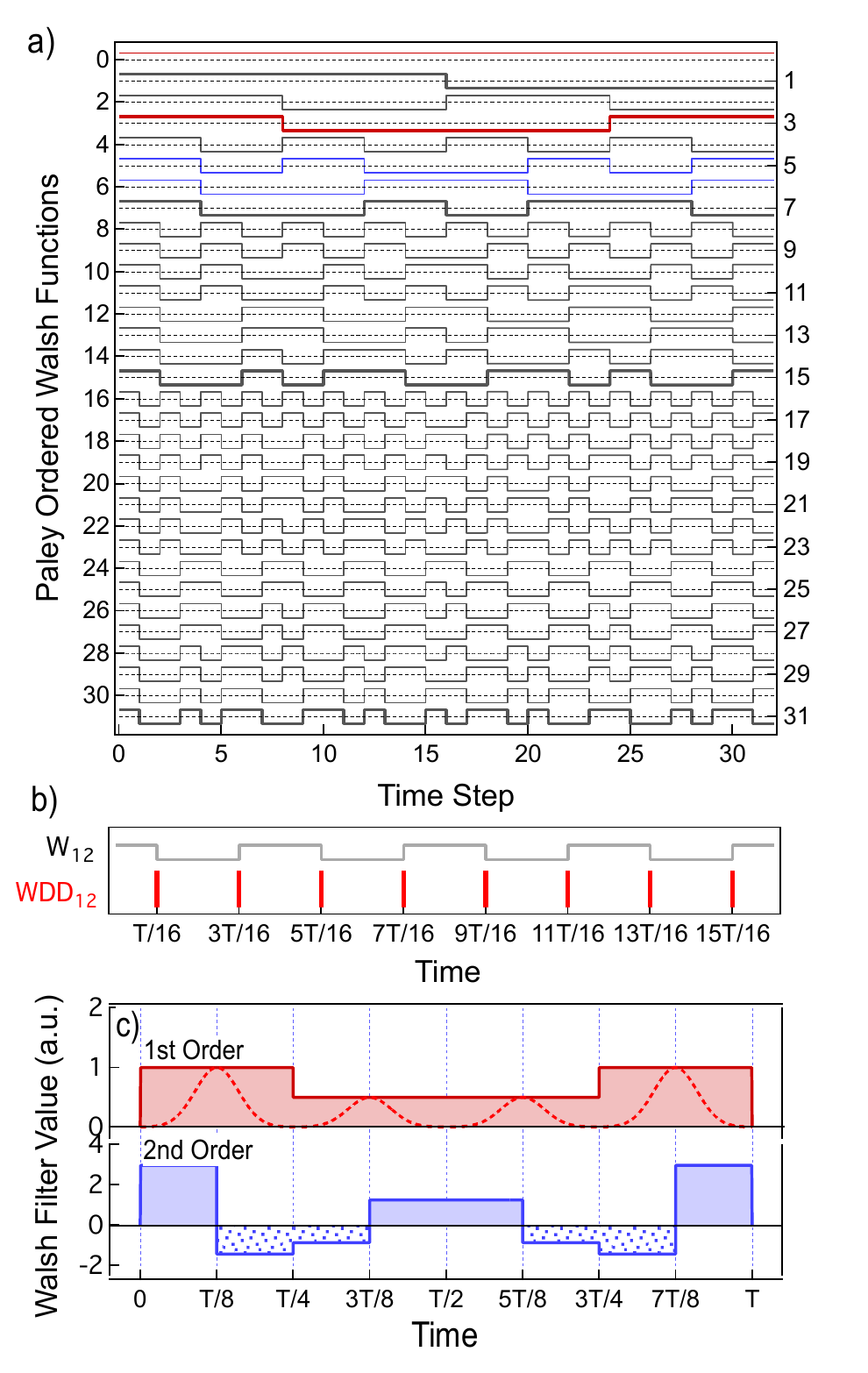}
      \caption{a) A finite set of the Walsh functions appearing in Paley ordering with normalized total durations.  Walsh functions highlighted in bold correspond to the \emph{Thue-Morse} subset, possessing Paley-ordered indices with the highest binary Hamming weight.  b) Example trace showing how Paley-order Walsh function with index 12, $\textsf{W}_{12}$ yields the pulse locations for a WDD$_{12}$ sequence via the location of digital transitions. Vertical lines schematically indicate control-pulse locations, normalized to total time, $T$. c) Example Walsh synthesized filter waveforms providing first and second order noise filtering.  Red trace composed of Walsh functions highlighted in red above, blue trace composed of Walsh functions highlighted in both red and blue.  Red dashed line indicates that arbitrary shapes for individual pulse segments can be accommodated with the Walsh waveform defining the overall amplitudes of each segment~\cite{Ball2014}.
}
      \label{Fig:Walsh}
\end{figure}

\emph{Use for a diversity of applications in quantum control}: This includes dynamical decoupling for quantum memory (Fig.~\ref{Fig:Walsh}b) ~\cite{HayesPRA2011, LongStorage}, Walsh-modulated (Fig.~\ref{Fig:Walsh}c) dynamically protected single-qubit gates (including well known composite pulses)~\cite{Ball2014, SoareNatPhys2014, True}, dynamically protected multiqubit gates~\cite{HayesPRL2014, GreenPRL2015}, noise spectroscopy and real-time quantum signal reconstruction~\cite{Cooper14}, the construction of quantum algorithms~\cite{Leung2002} (language therein refers to Hadamard matrices), and the construction of quantum circuits for the implementation of unitary operators~\cite{AAGuzik_2014, Svore2014}.  Across various applications, the Walsh family of controls is shown to incorporate previously known sequences and protocols in a single unifying mathematical framework.

\emph{Compatibility with classical digital hardware}: This arises from their bi-valued representation and composition from segments with minimum duration $\tau=\beta T_{c}$, defined as integer multiples of an underlying digital clock cycle, $T_{c}$.  The period $\tau$ defines all other periods within a Walsh function, and gives total sequence duration as a power-of-two multiple of $\tau$. This structure is compatible with both minimum-timing~\cite{Hodgson2010} and timing-precision~\cite{BiercukJPB2011} constraints in DES, and is also easily realized in hardware (see Sec.~\ref{Sec:Output}).

\emph{Simple programming}:  The error suppressing properties of Walsh DES protocols are captured via a small number of critical parameters, such as the Paley-ordered index of the selected Walsh function, and Walsh-synthesized DES waveforms are sparse in the Walsh basis.   For instance the red trace appearing in Fig.~\ref{Fig:Walsh}c consists only of a weighted sum of Walsh functions $\textsf{W}_{0}$ and $\textsf{W}_{3}$ (subscripts denote the Paley-ordered index, see Sec.~\ref{Sec:Output}), but still provides first-order noise filtering comparable with more complex modulation techniques~\cite{Ball2014, SoareNatPhys2014}.   As a result only limited information is required to define a high-performance Walsh-DES protocol, minimizing the classical communication necessary to program the controller as well as the memory required to produce and store a modulation waveform.  

These characteristics, and the demonstrations surveyed in Appendix~\ref{Ap:WalshOverview}, provide compelling evidence that Walsh functions form a suitable programming basis for the implementation of the physical layer controller.  In summary, if we restrict ourselves to timing patterns and controls readily represented in mathematical framework of Walsh functions, we will be able to both accomplish a wide range of relevant control tasks and also streamline the architecture of the control hardware.  We therefore proceed by considering how the key functionality required of a physical layer controller may be mapped to the design and programming of hardware using the Walsh basis.

\section{Experimental implementation of an FPGA Walsh controller}\label{Sec:FPGA}  
 In order to demonstrate the value of this design framework for physical layer classical controllers we are motivated to develop a real-time Walsh controller based on a field-programmable gate array (FPGA).  Synthesizing the practical demands associated with the application of quantum control operations in the Walsh framework for either algorithmic sequencing at the physical level or DES, we define a set of summary requirements which capture the required output performance and characteristics of the controller.  These requirements then inform the specific experimental demonstrations which follow.   A broad overview of the use of Walsh functions in quantum control providing  justification of these requirements is presented in Appendix~\ref{Ap:WalshOverview}.  For full mathematical details describing the Walsh functions in the context of DES see Refs.~\cite{HayesPRA2011, Ball2014}. \\
 \\
\noindent \emph{Summary Requirements:}
\begin{enumerate}[(1)]
\item Generate time-scaled Walsh functions to establish timing patterns for algorithmic execution or DES application. 
\item Synthesize analog waveforms or digital programming instructions from weighted sums of Walsh functions and use them to modulate a control system in time. 
\item Trigger the generation and output of an analog Walsh-synthesized waveform or other system output synchronously with the Walsh timing system.
\item Repeat the resulting Walsh-timed sequence of operations a user-defined number of times, meeting needs for long-time memory storage.
\item Employ Walsh-based signal reconstruction and spectrum analysis using sensor inputs derived from e.g. dedicated sensor qubits.
\end{enumerate}


Even with these functional requirements in hand, there are many approaches to realizing appropriately synthesized outputs in an embedded FPGA controller, and we must therefore consider system-level design drivers that will narrow our choice of implementation.
\\
\\
\noindent \emph{Design Drivers:}
\begin{enumerate}[(1)]
\item The need for ``lightweight'' designs minimizing the use of computational hardware in order to maximize scalability.  

\item Constraints on communication bandwidth to the embedded controller, mandating simple programming, especially when deployed in cryogenic environments.

\item The ability to adjust controller outputs with minimum latency based on new information derived at the physical level or from higher-level algorithmic instructions 
\end{enumerate}

Consideration of these design drivers narrows the space of acceptable controller designs by steering implementation towards real-time waveform synthesis instead of precompilation of output control waveforms within the embedded controller.  This approach dramatically reduces hardware memory requirements and also minimizes the latency to output of the desired waveform.  We proceed by providing an example implementation meeting both the Summary Requirements and the Design Drivers articulated above.  The architecture which we employ does not represent a unique solution, but as we will show, achieves the desired performance with significant gains relative to alternate approaches.

In the remainder of this section we describe and experimentally demonstrate the functionality achieved for our Walsh-programmed controller, and provide a detailed description of the hardware implementation.  We broadly divide discussion of FPGA performance into the tasks of control-signal generation and system identification.  We begin our discussion by introducing the Walsh Generator as the fundamental module used throughout this work, and then describe how various subsystems function in combination to permit a broad range of complex outputs within the Walsh basis.

\begin{figure}[tp]
      \centering
      \includegraphics[width=0.9\columnwidth]{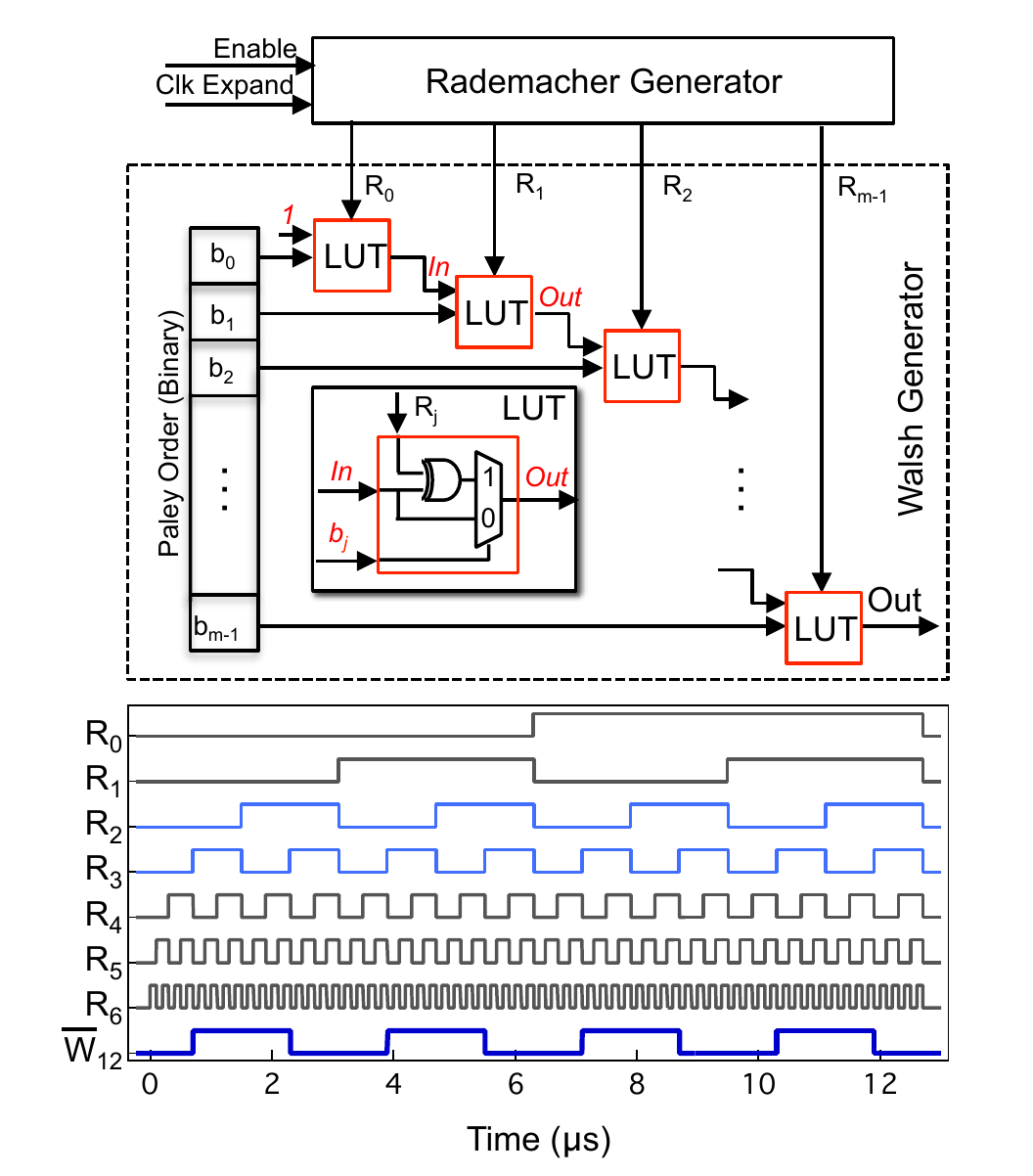}
      \caption{Schematic of a high-efficiency real-time Walsh-function generator implemented in FPGA. Square-wave Rademacher functions are summed with the binary digits of the desired Walsh function in Paley ordering via a cascade of lookup tables (LUT), each composed of an XOR gate and a multiplexer.  Timing of the Rademacher functions is set by the Clk Expand input such that the base time period of the Walsh function is a fixed multiple of the clock period. The Enable input begins the signal generation via an external hardware trigger.  Lower panel shows real-time oscilloscope data-capture of Rademacher functions of various orders, with period increasing by a factor of two with each increase in the Rademacher order.  The two light blue Rademachers ($\textsf{R}_{2}$ and $\textsf{R}_{3}$) undergo modular addition via the Walsh Generator to form the output Walsh function with Paley order 12 ($12=2^{3}+2^{2}$, or $12\to 1100$, $b_{2}=b_{3}\neq0$), shown in dark blue at the bottom of the figure.  Additional circuitry has been constructed to route Rademacher functions to digital output pins for this diagnostic measurement.  In this demonstration we employ a Xilinx Zynq 7010 as part of a Dual-core ARM Coretex A9 + FPGA system-on-chip packaged by Red Pitaya.}
      \label{Fig:WalshGen}
\end{figure}

  \begin{figure*}[ht]
      \centering
      \includegraphics[width=15cm]{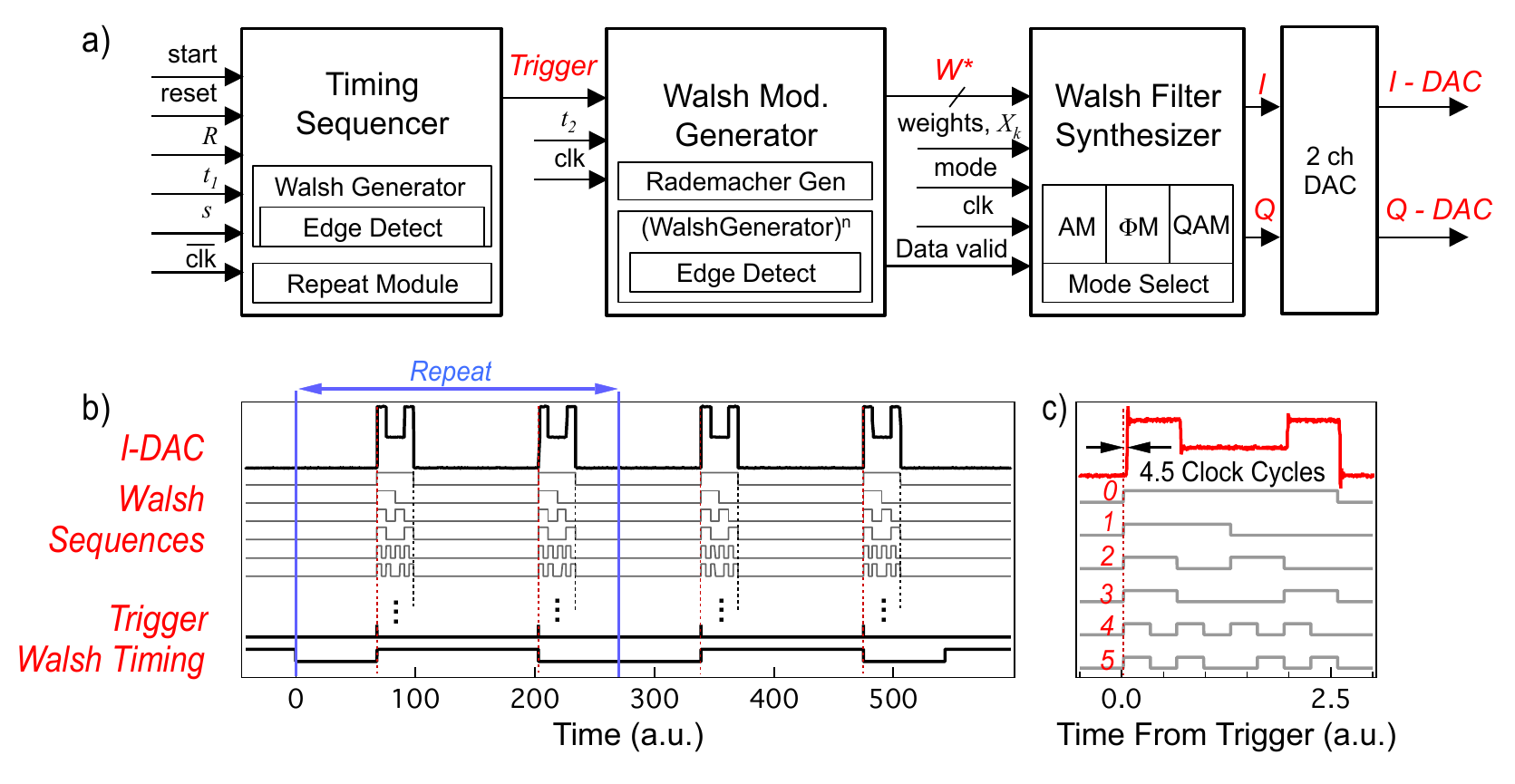}
      \caption{Schematic of a high-efficiency real-time Walsh-based controller implemented in FPGA.  a) Key subsystems used to fulfil the \emph{Summary Requirements} listed in the main text.   Inputs to the Timing Sequencer include ``start'' which triggers the output, ``reset'' which resets the state of the system, ``repeat'' which determines, $R$, the number of sequence repeats , ``$t_{1}$'' which sets the clock expansion on signal ``clk'' for the primary Walsh timing signal and ``order'' which selects the Paley ordered Walsh function to define the timing signal.  The Walsh Modulation Generator takes as input the ``Trigger'' signal output by the Timing Sequencer as well as another clock and clock-expansion parameter.  The Walsh Filter Synthesizer takes input Walsh sequences, denoted {\sl W*} as well as ``weights'' which define the analog Walsh synthesis coefficients, ``mode'' which selects the modulation protocol, and a clock. All clock signals are generated by a phase-lock-loop.  We employ two clocks with equivalent frequencies but differing by a $\pi$ phase shift for the Timing Sequencer and the two subsequent modules respectively; this results in a half-clock-cycle delay as observed in panel (c).   b) Representative example outputs of the relevant main subsystems using our FPGA implementation and captured using an oscilloscope.  Red vertical dotted lines indicate trigger locations associated with Walsh timing, here $\overline{\textsf{W}}_{3}$ repeated twice. In this example I-DAC corresponds to an AM-only Walsh noise filtering protocol highlighted in c) The modulation waveform for each pulse provides a first-order Walsh Amplitude Modulated Filter via a weighted sum of $\textsf{W}_{0}$ and $\textsf{W}_{3}$.  The first six Walsh functions generated by the Walsh Modulation Generator on a trigger from the Timing Sequencer are shown.  Again, in order to access signals not typically routed directly to FPGA outputs (e.g. the synthesized Walsh functions), additional FPGA circuitry has been implemented for diagnostics only.  }
            \label{Fig:WalshSchem}
\end{figure*}

\subsection{Architecture for controller output-stage}\label{Sec:Output}
 The performance of our FPGA controller architecture is summarized in Figs.~\ref{Fig:WalshGen} and \ref{Fig:WalshSchem}, building on modules described in detail in Appendix~\ref{Ap:Modules}.   The key element of our approach is a hardware-efficient Walsh Generator (Fig.~\ref{Fig:WalshGen}) performing modulo-2 addition of Rademacher functions~\cite{rademacher1922} according to the binary representation of the Paley-ordered Walsh index~\cite{Paley1932}, via combinatorial logic. We construct the Walsh function of Paley order $l$ as 
\begin{equation}
\overline{\textsf{W}}_{l}(x) =\overset{m-1}{\underset{j=0}{\text{\Large$\oplus$}}}{b_{j}\textsf{R}_{j}(x)}\label{Eq:WalshDef}
\end{equation}
\noindent where $x\in [0,1]$ defines the continuous domain, $b_{j}\in\{0,1\}$ are bit-values in the binary representation of $l$, written $(b_{m-1}b_{m-2}...b_{0})_{2}$, and $\oplus$ denotes modulo-2 addition. The building blocks in this construction are the Rademacher functions, with the $j$th order Rademacher defined by 
\begin{equation}\label{Eq:RademacherDefinition}
\textsf{R}_{j}(x)=[1-\textrm{sgn}\left[\sin\left(2^{j+1}\pi x\right)\right]]/2,\;\; j \geq 0.
\end{equation} 
\noindent This corresponds to a digital ``clock'' starting in the 0 state, and switching between state values $\{0,1\}$ with frequency $2^{j}$ (period $2^{-j}$).  Following Eq.~\ref{Eq:WalshDef}, target Walsh functions are generated in real time via composition of the Rademacher functions under modulo-2 addition. An example of this construction for Walsh function $\overline{\textsf{W}}_{12}(x)$ is presented in the lower panel of Fig.~\ref{Fig:WalshGen}. All Walsh functions defined by Eq. \ref{Eq:WalshDef} start with initial value 0, and in particular the constant-valued zeroth order function is $\overline{\textsf{W}}_{0}(x) = 0$. For some tasks, however, it is useful to redefine the Walsh basis with a non-zero value for the zeroth-order. For this we define the complement Walsh basis via a bit-inversion of Eq. \ref{Eq:WalshDef} as 
\begin{equation}
\textsf{W}_{l}(x) = \overline{\textsf{W}}_{l}(x)\oplus 1 \label{Eq:WalshDefComplement}
\end{equation}
which is also generated by the Walsh Generator module in Fig.~\ref{Fig:WalshGen} after a trivial modification. 

Leveraging the ability to generate Walsh functions of arbitrary order in real time we develop complex modules for Walsh-timing signal generation and Walsh-based waveform synthesis, in order to deliver on the Summary Requirements outlined above.  All submodules are synchronized to the same 100 MHz system clock supplied by the FPGA hardware.  First, as represented in Fig.~\ref{Fig:WalshSchem} the Walsh Timing Sequencer determines the timing of application of physical-layer control operations such as WDD, with control operations applied on rising and falling edges of the Walsh functions scaled temporally by parameter $t_{1}$. For this task we use Walsh functions of the form $\overline{\textsf{W}}_{l}(x)$ where the first rising edge occurs after $x = 0$, so it remains useful as a timing trigger in real devices with (small) finite physical delay times. Another critical task of the FPGA is to generate, in real time and on a designated trigger, a set of Walsh functions which are then used to form an analog waveform as a superposition via standard techniques in Walsh synthesis. Outputs of the Walsh Timing Sequencer generated by an Edge Detect submodule trigger a Walsh Modulation Generator module, which uses $n$ copies of the Walsh Generator to produce, in parallel, the first $n$ Walsh functions in Paley order.  For this task we use Walsh functions of the form $\textsf{W}_{l}(x)$ to enable waveform synthesis with a non-zero constant-offset component ($\textsf{W}_{0}(x) = 1$).

The resultant Walsh functions, scaled by timing information $t_{2}$, are fed into the Walsh Filter Synthesizer where they are combined in real time with user-defined weights, $X_{k}$.  The outputs of this module are streaming digital programming instructions representing a modulation waveform for a control operation within the Walsh family.   We customize solutions for Walsh filter synthesis for three key modulation techniques in use:  amplitude modulation (AM) of a control signal implementing a Pauli-$x$ operation for suppression of dephasing noise; phase modulation ($\Phi$M) for suppression of amplitude damping noise; and Quadrature Amplitude Modulation (QAM) employed for simultaneous suppression of noise in both quadratures.  Through this approach we can realize a wide variety of DES modulation protocols such as Walsh amplitude modulated filters~\cite{Ball2014}.  
 
Our controller architecture segregates pulsed-carrier synthesis (resonant with the qubit transition) from baseband modulation generation, providing a range of modulation implementation options.  For instance, we know that Walsh-synthesized modulation waveforms may be used either to directly define the shape of a square pulse, or to define the maximum-amplitude scaling of pulse segments with arbitrary shape (e.g. Gaussian), as shown in Fig.~\ref{Fig:Walsh}c.  Hence the downstream use of signals $I$ and $Q$ may be tailored to the needs of a specific hardware system (see Fig.~\ref{Fig:WalshSynth} in Appendix~\ref{Ap:Modules} for details).  In our example implementation these signals are constructed to directly program a digital-to-analog converter (DAC).  The DAC outputs voltages that may be used to drive the in-phase (I-DAC) and quadrature (Q-DAC) channels of an $IQ$ mixer which directly modulates the near-resonant carrier.   
 
Example digital and analog output signals captured by a mixed-signal oscilloscope are demonstrated in the lower panel of Fig.~\ref{Fig:WalshSchem}. Here the output is a first-order Walsh-Amplitude-Modulated Filter triggered synchronously with a repeated $\overline{\textsf{W}}_{3}$ timing pattern.  This particular gate has been experimentally demonstrated to provide efficient suppression of errors induced by non-commuting dephasing noise~\cite{SoareNatPhys2014}.  Timing, given by a repeated Walsh sequence, could be used for WDD application in error suppression or system identification as described in the next section. Total system latency from trigger to output of the analog waveform is only 4.5 clock cycles.  A detailed performance comparison relative to programming of a conventional microcontroller is provided in Sec.~\ref{Sec:Comparison}.



\subsection{Real-time Walsh signal reconstruction}\label{Sec:SID} 
Another set of tasks in implementing the embedded controller relates to the application of sensing and signal-reconstruction techniques using embedded sensors at the hardware level.  We proceed by defining an approach consistent with Walsh control and leveraging concepts developed in~\cite{Cooper14} for real-time signal reconstruction using Walsh-modulated quantum sensors.  The implementation of a real-time Walsh-signal-reconstruction protocol outlined in~\cite{Cooper14} requires functionality similar to that described above.  This primarily includes the ability to sequence WDD on multiple parallel sensors and then process the resulting projective measurements in order to determine a decomposition of the underlying noise process in the Walsh basis. 

Given an ensemble of qubits, each subjected to a time-varying dephasing field $b(t)$, and modulated by a Walsh dynamic decoupling sequence of different orders defined by $\textsf{W}_k(t/T)$, each will accumulate a net dynamic phase $\phi_{k}(T)$.  This phase is linked to a measurable quantity $P_{k}$ defining the weight of each Walsh function in a decomposition as \begin{align}\label{MainEq:Computed Walsh Weight}
X_k(T)  = 
\frac{\sin^{-1}[2P_k-1]}{\gamma T}.
\end{align}
\noindent With the resulting Walsh spectrum over order $k$ in hand one may reconstruct an approximation of the underlying noise signal as \begin{align}\label{MainEq:rec_walsh_cts}
\hat b(t) &= 
\sum_{k=0}^{N-1}  X_k \textsf{W}_k(t/T).
\end{align}
\noindent See Appendix~\ref{Ap:SID} for full details of the measurement protocol.

\begin{figure}[tp]
      \centering
      \includegraphics[width=0.9\columnwidth]{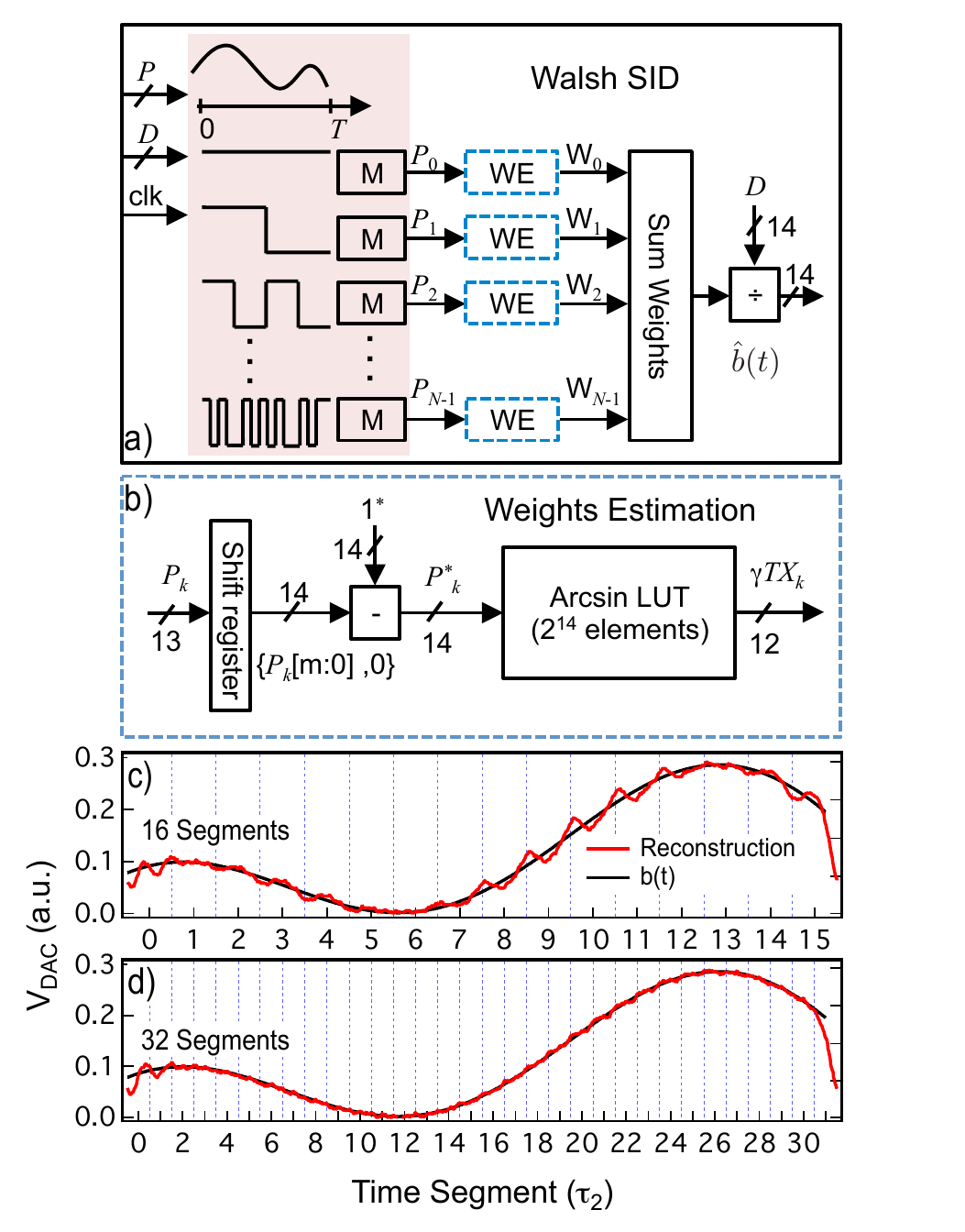}
      \caption{Schematic of Walsh-based systems identification (SID) module implemented in FPGA. a) Here the red shaded box indicates a protocol performed either at the qubit level, or as here, in simulation, returning fidelities $P_{k}$ from measurements $M$ and feeding them to Walsh Estimation modules ( b). The input parameter $D$ is responsible for scaling the the arcsin function, taking the value $D = \gamma T$. The output of weight summation module is the reconstructed signal $\hat{b}(t)$ for the magnetic field, represented digitally and here used to program a DAC for signal monitoring. b) Detailed schematic of Weights Estimation. The 13-bit binary presentation of Fidelity ($P_k$) doubles its value by using a shift-left 1-bit register and then subtracts to 14-bit binary presentation of ``1'' to derive $\arcsin(\gamma TX_k)$ as shown in Eq. \ref{Computed Walsh Weight}. The arcsin LUT is a $2^{14}$-element look-up table which returns $TX_k$. A divider is used to scale the output of arcsin LUT  with divisor $T$ and output the final result Walsh weights $X_k$. The selected value of the bit-depth for the arcsin function is determined by the specific hardware implementation employed here. c)-d) Analog waveforms output from the DAC overlaid with applied noise signal given numerically calculated values for $P_{k}$, using 16 or 32 Walsh functions (hence, individual fidelity measurements).}    \label{Fig:WalshSID}
\end{figure}

With this framework in mind, we design a Walsh SID module to calculate Walsh weights, and produce real-time reconstructions $\hat{b}(t)$ from a given set of fidelity values. The input information for this module is the vector of fidelity values, $\boldsymbol{P} = (P_0, P_1,...,P_{N-1})$ generically describing the measurement data of the sensor array, though here generated from numerical simulations (Appendix \ref{Ap:SID}). These values are converted into estimates of the corresponding Walsh weights using Eq. \ref{MainEq:Computed Walsh Weight}, and subsequently incorporated in a Walsh waveform synthesizer module to implement a real-time reconstruction $\hat{b}(t)$ on an FPGA. The central capability of the Walsh SID module is captured in the Weights Estimation module shown in Fig.~\ref{Fig:WalshSID}, with reconstruction of an analog waveform subsequently produced by the Sum Weights module relying on the same architecture as the Walsh Filter Synthesizer described in Sec. \ref{Sec:WalshSynthesisForControlWaveforms}.

\subsection{Comparative performance of the Walsh controller}\label{Sec:Comparison}

The key strengths of this approach are the ability to produce outputs in real time with limited hardware resources dedicated to the controller.  For instance on the Xilinx ZC7Z010 FPGA device, one of the smallest in the System-on-Chip family, implementation of the scheme in Fig.~\ref{Fig:WalshSchem} for the first 8 Paley orders, occupies less than 5\% of the available 17,600 lookup tables (LUTs). Moreover, devices such as the XC7Z100 have 277,400 LUTs, this being a factor of 15 larger. The direct mapping of Walsh functions to simple digital logic constructions allows for a highly autonomous system with minimal requirements on communication bandwidth to the controller, requiring only 345 bits for full programming.  This efficiency arises because our control solutions are sparse in the Walsh basis.  

A key performance metric for this system is hardware latency, which is optimized through use of a hardware-timing solution.  In our implementation, from trigger input to the start of signal generation requires only 4.5 clock cycles and from changes in input variables to ready-to-output status requires only two clock cycles.  The system may be reset at any time and returns to ready-to-output status in three clock cycles.  Key contributions to the output delay are as follows:

For comparison we built a system with similar functionality using an embedded microcontroller (Arduino MEGA 2560).  In this implementation, because the output is compiled rather than generated by hardware in real time, the entire output waveform must be calculated and stored in memory before the system is ready-to-output.  This process requires nearly 1.3 M clock cycles for the waveforms demonstrated above, and must be performed every time a variable is changed or reset signal received.  Similarly, Walsh-sequence calculation is 8,000 times slower on the microprocessor.  Other challenges with timing resolution, synchronization, and repeatability due to the microcontroller interpreting instructions to produce output signals make this implementation far inferior relative to the FPGA.  For instance, the sequential architecture of the microcontroller requires that instructions be completed before the system responds to a change in state; this results in delays from trigger to output of an AM envelope of $\sim125$ clock cycles, contrasted to just 4.5 in the FPGA.  Similarly, any update to variables such as the Walsh order, Walsh weighting coefficients, or timing information requires full calculation of the waveform, or $\sim1.3$ M clock cycles.  Importantly, these delays would persist in the microcontroller solution regardless of the selected functional basis for waveform creation. 

 Finally, because many high-performance modulation patterns are sparse in the Walsh basis (e.g. the example in Fig.~\ref{Fig:Walsh}c consisting only of a weighted sum of $\textsf{W}_{0}$ and $\textsf{W}_{3}$) they may be generated in real time using simple digital logic on the FPGA, in contrast to generating a similar modulation waveform using e.g. Fourier synthesis in either a microcontroller or an alternate FPGA-based synthesizer.  For instance, assuming a base period $\tau$ and clock cycle, $T_{c}$, generating the same waveform using Fourier synthesis would require at least $\tau/T_{c}$ sine functions to be combined to achieve the same timing resolution.  Such an approach would lead to a significant expansion of required memory allocation, programming complexity, and combinatorial logic. 

The Walsh SID module has similar performance characteristics of reset-to-ready delay time or changes to ready-to-output latency, with trigger-to-output requiring 5 clock cycles to produce a valid output. This performance metric plays an important role when we try to optimize the latency of each design module to achieve a real-time response.  Low output latency requires significant hardware resource allocation, with particular focus on the arcsin LUT, which consumes $\sim5\%$ of Block Ram (total 60 units) for a single instance. Other modules remain resource efficient, maintaining hardware requirements at less than 1\% for FF (Flip Flops) or 6\% of LUTs (total 35,200 FFs and 17,600 LUTs). The trigger-to-output latency distribution are listed as follows: The Weights Estimation module generates the parameter $\gamma TX_k$ in two clock cycles. The output of this module will enable the Sum Weights module (Fig. \ref{Fig:WalshSID}) to begin reconstruction of the digital representation of $\hat{b}(t)$ after two clock cycles.  The final divider module scales the output amplitude within one clock cycle. 





\section{Conclusion}\label{Sec:Conclusion} 
In this work we have explored the functional drivers and system-level constraints associated with implementing physical-layer control, and used them to identify the Walsh functions as a useful functional basis for the programming and design of efficient embedded classical controllers.  We have considered the requirements of dynamic error suppression and physical-layer operation sequencing in order to develop a notional design for a Walsh-based controller.  Following this we have implemented this architecture on an FPGA and demonstrated considerable performance advantages relative to approaches leveraging general-purpose signal synthesis via microcontrollers.  Alternative FPGA architectures may be considered, but we emphasize that our approach of real-time Walsh-signal synthesis and SID provides substantial benefits in terms of FPGA resources and output latency -- both key drivers in the context of designing embedded physical layer controllers in quantum information.

Overall we believe that the requirements of implementing physical-layer control, and in particular physical-layer error evasion, have largely been overlooked in past architectural studies in favor of mapping physical qubit layouts to QEC-code connectivity.  This is understandable given the extraordinary demands arising from consideration of QEC, but is in our view insufficient to consider realistic quantum information architectures at scales where logical qubit error rates will need to be exceedingly small (comparable to the inverse problem size~\cite{Kubiatowicz}).  By showing how existing analyses may be augmented through consideration of the design of embedded physical-layer controllers, this work advances broad architectural studies in a manner consistent with our current understanding of projected physical-layer error-rates and control requirements. In addition, this work addresses the general question of how to efficiently sequence and apply physical-qubit manipulations that are required even without use of DES.   Future work will consider the integration of FPGA Walsh controllers in quantum information experiments, and detailed consideration of the interface between the classical controller and logical layers~\cite{JonesPRX2012}.

\begin{acknowledgments}
The authors acknowledge useful conversations with D. J. Reilly, W. D. Oliver, and L. Viola.  Work was partially supported by the ARC Centre of Excellence for Engineered Quantum Systems CE110001013, the US Army Research Office under Contract W911NF-12-R-0012, the US ODNI via the IARPA LogiQ program, the Australian Institute for Nanoscale Science and Technology (AINST) Accelerator Scheme, and a private grant from H. \& A. Harley. 
\end{acknowledgments}

\newpage
\bibliography{STIQSBIB} 

\appendix

\section{Physical layer controls in the Walsh basis}\label{Ap:WalshOverview}
In the following subsections we provide greater detail on the specific application of Walsh functions in the implementation of physical-layer quantum control.  

\subsection{Walsh timing in quantum control}\label{Sec:Timing}
Walsh sequences can be used to provide timing information relevant to the sequencing of control operations on qubits at the physical level, with rising and falling edges employed to trigger the application of a desired operation.  Generically this can be employed for logical or physical-layer algorithmic implementation or unitary gate construction~\cite{Leung2002, AAGuzik_2014, Svore2014}, and of course it may also be advantageous to apply all physical-layer operations timed via Walsh sequences due to their discrete timing structure. We therefore refer generically to Walsh \emph{timing} patterns.

The simplest application of Walsh functions in the algorithmic application of DES specifically comes through association of the Walsh function itself with the control propagator of a dynamical decoupling (DD) sequence for implementation of the identity (memory)~\cite{HayesPRA2011}.  Each digital transition in a Walsh function corresponds to the location of a control pulse in the resulting WDD sequence, Fig.~\ref{Fig:Walsh}b.  Incredibly, a number of familiar DD sequences can be identified as special instances of WDD sequences:  Using the Paley-ordering convention as an index $\text{WDD}_{0}$ (a WDD sequence with pulses prescribed at the transitions in a $\textsf{W}_{0}$ Walsh function) is free evolution and $\text{WDD}_{1}$ is the spin echo sequence. Similarly, special instances of Walsh dynamical decoupling correspond to Periodic Dynamical Decoupling (PDD), CPMG, and Concatenated Dynamical Decoupling~\cite{HayesPRA2011}.  Changing the value of $\tau$, the base period of the Walsh function, changes the frequency response of the WDD sequence~\cite{GreenNJP2013}. Importantly, the order of error suppression realized in a WDD sequence is determined entirely by a single number, $r$, the Hamming weight (number of ones) of the binary representation of the Walsh function's Paley-ordered index.  

For generalized storage of quantum states we previously identified that it is possible to reconcile the oppositional demands for high-fidelity state storage and low access latency to stored states protected by DD via periodic repetition of a base, high-order DD cycle~\cite{LongStorage}.  This insight led to demonstrations that repetition of WDD sequences could result in conditions by which a target memory fidelity could be guaranteed at long storage times, but with access latency capped at the duration of the base WDD cycle.  Moreover repetition is a natural mathematical operation within the Walsh family, meaning that repeated sequences generally also fall within the Walsh family.

\subsection{Walsh-synthesized modulation for noise-suppressing control operations}

The implementation of DES single and two-qubit operations requires the application of temporally modulated control fields (e.g. frequency, phase, or amplitude of a near-resonant carrier) which enact the qubit manipulations with added robustness against environmental or control fluctuations.  Following the general approach of composite pulsing we consider an arbitrary operation decomposed into discrete time-segments; these modulated pulses would then be applied according to a precise timing pattern in order to implement a desired higher-level algorithm or protocol.

The time-domain modulation waveform defining the envelope for the phase or amplitude of a near resonant carrier may be crafted using Walsh synthesis, known to facilitate construction of arbitrary waveforms via superpositions of Walsh functions. The mathematical properties of the Walsh functions establish key analytic design rules simplifying the construction of high-performance Walsh modulation protocols for arbitrary single-qubit gates~\cite{SoareNatPhys2014,Ball2014}.  These protocols also bring the benefit that high-performance noise-suppressing solutions are sparse in the Walsh basis, requiring superposition of only a few Walsh functions (Fig.~\ref{Fig:Walsh}c).  In addition, a variety of well known dynamically corrected gates~\cite{Khodjasteh2010, Khodjasteh2009, aDCG} and NMR-inspired composite pulses~\cite{Brown2004, True} are included in the Walsh synthesis framework ( Ref.~\cite{Ball2014}).  

Two-qubit gates implemented via an intermediary bosonic mode -- such as geometric phase gates in trapped ions~\cite{Leibfried2003} and superconducting qubits~\cite{Schoelkopf2007} -- can be modulated using Walsh functions in order to improve robustness against fluctuations in the parameters of the mediating driving field (e.g. amplitude and frequency).  This includes phase-modulation applied to gates mediated by a single bosonic mode~\cite{HayesPRL2014} with error suppressing properties described as in WDD.  In addition, application of a phase-modulation pattern described by the Thue-Morse subset of the Walsh functions (i.e. those with Paley order $k=2^{n}-1$ for $n$ an integer), provides a means to decouple multiple bosonic modes from the qubit state~\cite{GreenPRL2015}.  Here, repetition of the relevant Walsh modulation pattern provides a means to improve the robustness of the gate to time-dependent fluctuations. Remarkably, even the phase values applied during each segment in a Walsh-timed modulation pattern may be represented using the properties of Walsh functions.

\subsection{Walsh-based SID}
In any control setting it is required that we have both a model of the underlying physical system to be controlled (including its response to external stimuli) and up-to-date information about any disturbances that may need to be suppressed or compensated.  In control theoretic language the determination of this information is broadly referred to as ``System Identification''~\cite{Stengel}, overlapping the quantum physics concept of Hamiltonian Estimation when we are addressing the characteristics of the noise processes to be suppressed via physical-layer control.  Information gained through such protocols may then be employed to adjust or even optimize the selected physical layer controls.

Techniques for system identification in noisy quantum systems already exist and are widely employed in the community at the physical layer.  For instance, dynamical-decoupling-based spectral reconstruction generally involves execution of a set of repeated DD sequences followed by projective measurements~\cite{Alvarez_Spectroscopy, Oliver_PC, Norris_Spectroscopy}.  The relevant DD sequences fall within the WDD family, and are restricted to those subsequences which have a narrow spectral response in the Fourier domain~\cite{BiercukJPB2011, Ozeri2013}.  By appropriately modulating the temporal evolution of the system-environment interaction with DD, measurement outcomes on the system become maximally sensitive to the corresponding spectral content of the noise.  These measurements are followed by processing and reconstruction of data to determine the underlying noise power spectrum.   

In addition to ensemble-averaged statistical information about extrinsic noise processes, real-time estimation of the the dominant noise processes may be derived following protocols described by e.g.~\cite{Cooper14}.  Here parallel qubit sensors are subjected to synchronized noisy evolution, while being modulated by WDD sequences of different orders, returning multiple projective-measurement outcomes.  From these fidelity measurements the underlying time-varying signal can be reconstructed in each discrete timestep, $\tau$, using concepts from Walsh synthesis.  In the limit that these measurement results are not performed synchronously the same information may be used for spectral reconstruction.

\section{Detailed description of FPGA modules}\label{Ap:Modules}
In this Appendix we provide technical details of the architectural approach to realize the FPGA controller modules outlined in the main text.  Here we focus on the Walsh Timing Sequencer and the Walsh Synthersizer.

\subsection{Walsh Generator}
Following Eq.~\ref{Eq:WalshDef}, target Walsh functions are generated in real time via composition of the Rademacher functions under modulo-2 addition. In our framework this is performed by the Walsh Generator module shown in Fig.~ \ref{Fig:WalshGen} which stores the binary representation of a given Paley order $l$ in a dedicated register. Each bit is routed to a lookup table where it is combined with the relevant Rademacher, together with the additional input signal from the preceeding-order Rademacher.  The lookup table is comprised of an $\textsf{XOR}$ gate and a one-bit multiplexer which returns \emph{In} \textsf{XOR} $\textsf{R}_{j}$ if the relevant bit in $l$ is high, and returns \emph{In} if the bit is low. The cascaded lookup tables in the Walsh Generator combine the Rademacher Generator output with the desired $l$ to produce the target Walsh function. Signal propagation along this cascade forms the critical path of the design, demonstrating delays of less than one clock cycle.  An example of this construction for Walsh function $\overline{\textsf{W}}_{12}(x)$ is presented in the lower panel of Fig.~\ref{Fig:WalshGen}. Walsh functions of the complement form given in Eq. \ref{Eq:WalshDefComplement} are similarly generated after a trivial modification. 

In hardware, Rademacher functions $\textsf{R}_{j}(x)$ are generated using the Rademacher Generator module shown in Fig.~ \ref{Fig:WalshGen}. This employs a counter to synchronize the change of state at integer multiples of the underlying clock's base period, $T_{c}$. The change-of-state is performed in real time using a comparator after the required number of clock cycles have elapsed. The output is therefore effectively a square wave with period given by ``expanding'' that of the underlying clock.

\subsection{Walsh Timing Sequencer}\label{Sec:WalshTiming}

 The Walsh Timing Sequencer module shown in Fig.~\ref{Fig:WalshSchem} outputs timing information in the form a Walsh function generated by an embedded Walsh Generator module  (\emph{Summary Requirement 2}). We denote this ``sequencing'' Walsh function by $\overline{\textsf{W}}^{(1)}_{s}(x)$ where $s$ is the user-defined Paley order, which in our implementation can be any positive 8-bit integer $s\le 255$. The subscript (1) indicates that this Walsh function serves the dedicated purpose of synchronization to distinguish from the application of Walsh functions for waveform synthesis described in the next section. 

Temporal scaling of the Walsh function is programmed by the input scaling factor $t_{1}$.  In the standard Walsh basis with normalized time domain $[0,1]$ the Walsh function of Payley order $s$ may be divided into $2^{m(s)}$ piecewise-constant segments of equal duration, where $m(s)$ is the bit width of $s$. Consequently $2^{-m(s)}$ is the minimum (piecewise-constant) segment duration. In our implementation, the time domain is rescaled such that the minimum segment duration becomes
\begin{align}
\tau_{1} \equiv t_{1} T_{c} 
\end{align}
where $t_{1}$ is an 8-bit integer and $T_c = 10$ ns is the clock cycle time set by the FPGA hardware. The time domain of the rescaled Walsh basis is therefore given by $[0,t_{1} T_{c}2^{m(s)} ]$ with the minimum timing resolution set by the clock cycle time $T_{c}$. Moving from continuous to discrete time, we drop the argument in our notation in Eq. \ref{Eq:WalshDef},  and the digital representation takes the form
\begin{align}\label{Eq:WalshTimingSequence}
\overline{\textsf{W}}^{(1)}_{s} &= 
\left[
\begin{array}{c}
\BinaryVecTS_{s1},
\BinaryVecTS_{s2},
\dots,
\BinaryVecTS_{s,2^{m(s)}}
\end{array}
\right]
\end{align}
where  the $i$th piecewise-constant segment of duration $\tau_{1}$ is represented by the subsequence  
\begin{align}
\BinaryVecTS_{si} &\equiv
\begingroup
  \renewcommand*{\arraystretch}{1.15}%
  \kbordermatrix{
 & 1            &  2          & \hdots           &  t_{1}    \cr
   &     \BinaryTS_{si},             & \BinaryTS_{si}       &\hdots       &\BinaryTS_{si} \cr
  }%
\endgroup,
\hspace{0.5cm}
\BinaryTS_{si}\in\{0,1\},
\end{align}
constructed from $t_1$ repetitions of the binary digit $\BinaryTS_{si}$ occurring once per clock cycle, implemented over the time interval $(i-1, i]t_{1} T_{c}$. The binary Walsh sequence described by Eq. \ref{Eq:WalshTimingSequence} is generated in real time as a stream of binary data by the embedded Walsh Generator module. 

We also include a Repeat Module (Fig.~\ref{Fig:WalshSchem}a) to extend the timing signal by sequential repetition of the basic timing sequence $\overline{\textsf{W}}^{(1)}_{s}$ an integer number of times $R$, yielding the concatenated signal denoted $[\overline{\textsf{W}}^{(1)}_{s}]_{R}$. In our framework $R$ is specified by a 4-bit integer, supporting up to 15 repetitions, however this be extended as required. The number of repeats is monitored by a counter and comparator, and controlled by an enable signal which is high when the counter value is less than or equal to $R$ and low for values beyond $R$. For $R=0$, the system output is disabled.  

The output of this module serves as a trigger signal to synchronize control operations with the transitioning edges (bit-flip events) in the concatenated signal $[\overline{\textsf{W}}^{(1)}_{s}]_{R}$. This is achieved using the Edge Detect module (Fig. \ref{Fig:WalshSchem}), which uses a D-type flip flop (DFF) configured to register bit-flip events in $[\overline{\textsf{W}}^{(1)}_{s}]_{R}$ coinciding with the falling edge of the system clock.

\subsection{Walsh synthesis for control modulation waveforms}\label{Sec:WalshSynthesisForControlWaveforms}

\noindent Another critical task of the FPGA is to generate, in real time and on a designated trigger, a set of Walsh functions which are then used to form an analog waveform as a superposition via standard techniques in Walsh synthesis. Through this approach we can realize a wide variety of DES modulation protocols~\cite{Ball2014}, where for efficiency we have implicitly segregated the task of carrier synthesis (resonant with the qubit transition) from baseband modulation generation. 

Using the same general architecture outlined above, the Walsh Modulation Generator module (Fig.~\ref{Fig:WalshSchem}) addresses this task by generating the first $n$ Paley-ordered Walsh functions in parallel to serve as a truncated basis for Walsh synthesis.  Our implementation accommodates a basis consisting of Paley orders specified by 8-bit integers, namely $n\le 255$. This basis is scaled in time such that the minimum segment duration is given by $\tau_{2} \equiv t_{2} T_{c}$ where the input scaling factor $t_{2}$ is a 4-bit integer. The time-domain therefore spans the duration $[0,t_{2} T_{c}2^{m(n)} ]$ where $m(n)$ is the bit width of $n$. All basis functions are generated simultaneously using the same Rademacher Generator and $n$ independent instances of the Walsh Generator module, each streaming binary sequences in real time over $n$ parallel channels. All $n$ Walsh Generators are triggered simultaneously (after a delay of one clock cycle), using a DFF. The completion time for all $n$ Walsh sequences is tracked by an auxiliary bit in the counter of the associated Rademacher Generator module. 

The $k$th Walsh function output of this module is represented by the digital sequence
\begin{align}\label{MainEq:DigitalWalshSequence_index_k}
\textsf{W}^{(2)}_{k} &= 
\left[
\begin{array}{c}
\BinaryVecTS_{k1},
\BinaryVecTS_{k2},
\dots,
\BinaryVecTS_{k,2^{m(n)}}
\end{array}
\right]
\end{align}
where  the $j$th piecewise-constant segment of duration $\tau_{2}$ is represented by the subsequence  
\begin{align}
\BinaryVecTS_{kj} &\equiv
\begingroup
  \renewcommand*{\arraystretch}{1.15}%
  \kbordermatrix{
 & 1            &  2          & \hdots           &  t_{2}    \cr
   &     \BinaryTS_{kj},             & \BinaryTS_{kj}       &\hdots       &\BinaryTS_{kj} \cr
  }%
\endgroup,
\hspace{0.5cm}
\BinaryTS_{kj}\in\{0,1\},
\end{align}
constructed from $t_2$ repetitions of the binary digit $\BinaryTS_{kj}$ occurring once per clock cycle, implemented over the time interval $(i-1, i]t_{2} T_{c}$. The superscript (2) indicates that these basis functions serve the purpose of modulation-waveform synthesis. 

\begin{figure}[tp]
      \centering
      \includegraphics[width=0.75\columnwidth]{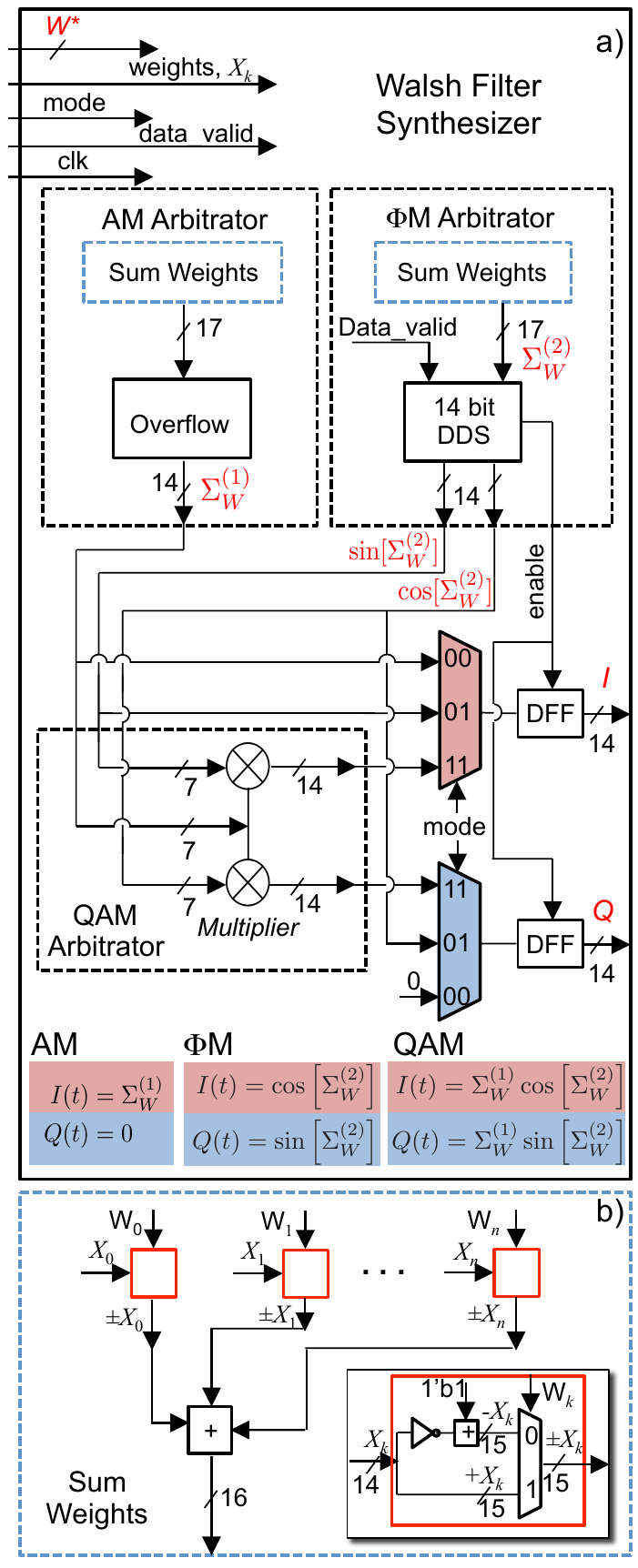}
      \caption{Schematic of the Walsh Filter Synthesizer employed in the Walsh controller.  bit widths of relevant signals are indicated.  AM, $\Phi$M, or QAM modulation are selected by the mode input, activating the relevant subunits.  We distinguish the outputs of the AM and $\Phi$M arbitrators using superscripts on $\Sigma^{(1,2)}_{W}$.  The $\Phi$M subsystems relies on in-built DDS blocks in the FPGA to directly rotate outputs to the appropriate basis.  The QAM Arbitrator takes 14-bit inputs and uses only the seven most-significant bits prior to multiplication.    This DDS is enabled by the ``data valid'' input generated by the Walsh Modulation Generator (Fig.~\ref{Fig:WalshSchem}).  Relationships of outputs under various modulation protocols, indicated by the mode input, are indicated at the bottom of the panel.  b) Within the AM and $\Phi$M modules we construct the Sum Weights submodule (blue dashed box).  Red boxes are highlighted in the inset and demonstrate the bitwise selection of $\pm X_{k}$ depending on the value of $\textsf{W}^{(2)}_{k}$.}
      \label{Fig:WalshSynth}
\end{figure}

These Walsh functions must now be combined in real time in the Walsh Filter Synthesizer with user-defined weights to produce waveforms appropriate for enacting a modulation protocol on the near-resonant carrier.   This superposition is represented as 
\begin{align}
\Sigma_{W} &\equiv
\sum_{k=0}^{n-1}[
X_k \times \textsf{W}^{(2)}_k
], \quad \textsf{W}^{(2)}_k \in \{\pm 1\}
\end{align}
where $X_k$ is weight of the \emph{k}th Walsh sequence, and the quantity $\Sigma_{W}$ is implicitly defined to be time-dependent and piecewise constant. The bit width of these weights is 14 bits, selected for compatibility with the underlying hardware employed for this demonstration, and includes a single bit to encode the sign of $X_k$ and the remainder for encoding the amplitude.   

This superposition is executed in the Sum Weights submodule illustrated in Fig.~\ref{Fig:WalshSynth}, in which each weight, $X_{k}$ is converted to a signed value $\pm X_{k}$ depending on the real-time state of $\textsf{W}^{(2)}_{k}$ using a multiplexer.   The output bit width is $\lceil14 + \log_{2}(n)\rceil$, where $n$ is highest Paley-order used in the Walsh synthesis protocol.  These digital outputs are then added together, checked against an overflow value, and sent to other subsystems. 

We customize solutions for Walsh filter synthesis for the three key modulation techniques in use:  amplitude modulation (AM) of a control signal implementing a Pauli-$x$ operation for suppression of dephasing noise; phase modulation ($\Phi$M) for suppression of amplitude damping noise; and Quadrature Amplitude Modulation (QAM) employed for simultaneous suppression of noise in both quadratures.  A 2-bit signal is used to indicate the mode of modulation (Fig.~\ref{Fig:WalshSynth}).

In AM mode, Walsh functions corresponding to negative weights $X_{k}$ are inverted.  The summation of all converted weights are linked to the input of the mode selector before being registered by DFFs to feed downstream hardware. 

In $\Phi$M mode, the modulated control signal may be conveniently represented in the $IQ$ basis through the following form:
\begin{align}
m_{\Phi M}(t) &= 
\cos\left[\Sigma_{W}\right]\cos(\omega t )-\sin\left[\Sigma_{W}\right]\sin(\omega t )\\
&= I(t)\cos(\omega t ) - Q(t)\sin(\omega t )
\end{align}
where $\Sigma_{W}$ represents the required phase as a function of time, and $\omega$ is the frequency of the carrier signal.  A dedicated arbitrator calculates the total phase shift $\Sigma_{W}$, and we then convert to the $IQ$ basis.  This process is accomplished in hardware by exploiting the trigonometric form of the in-phase and quadrature modulation coefficients, using an in-built Direct Digital Synthesis (DDS) module to function as a look-up table with 13(14)-bit input (output) (one-bit for sign, 13 for amplitude).  Here, a ``data valid'' signal is employed as a trigger derived from the composite architecture as detailed in the following section.  The DDS also outputs an ``enable '' signal controlling DFFs to prevent glitches due to a different clock period between the DDS and FPGA core.  From trigger to output requires one clock cycle in the DDS.  

The AM and $\Phi$M arbitrators are used synchronously for QAM, with outputs multiplied within a dedicated QAM arbitrator.   DFFs are added to account for a one-clock-cycle delay between the AM and $\Phi$M arbitrator outputs, and an output data valid signal from DDS is used to enable the latching of output to suppress glitches.   The use of multipliers in the QAM arbitrator adds another two-clock-cycle delay to the output when this mode is selected.  

\subsection{Clocking}
\begin{figure}[bp]
      \centering
      \includegraphics[width=0.75\columnwidth]{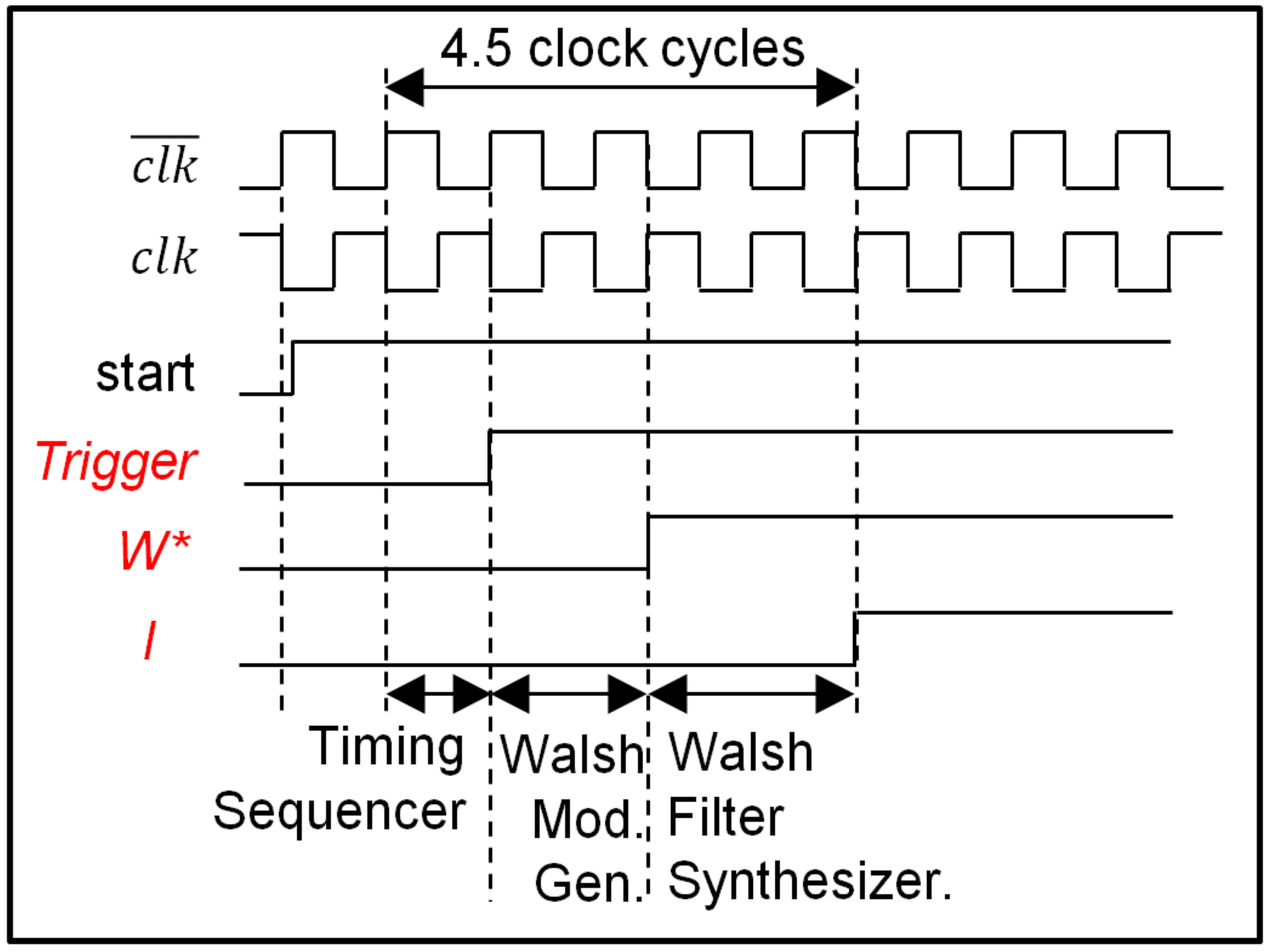}
      \caption{Waveforms illustrating timing and latency distribution for generating output signals via the FPGA modules. 
Clock signals (first two waveforms) are generated by a phase lock loop to minimise clock skew. These generated clocks take one of two forms: ``clk'' (used for the Timing Sequencer) or the complement form ``$\overline{\text{clk}}$'' (used for the Walsh Mod. Generator \& Walsh Filter Synthesizer). Latency distribution of internal signals ``start'', ``trigger'', ``$W^*$'' and output signal ``I'' are also indicated in terms of the number of clock cycles (see Fig. \ref{Fig:WalshSchem}).}
      \label{Fig:WalshLatency}
\end{figure}

To demonstrate how we may reduce latency between the output of the Timing Sequencer and the Walsh Modulation Generator our architecture employs two distinct clock signals. These are generated with minimal skew by a phase-locked loop.  As indicated in Fig ~\ref{Fig:WalshLatency}, the Timing Sequencer relies on a clock of the form designated by ``clk'', to drive an internal ``Trigger'' signal with one clock cycle delay after the start signal. The subsequent modules use clock signals of the complementary form $\overline{\text{clk}}$, phase shifted by $\pi$ relative to ``clk''. This results in a delay of 1.5 and 2 clock cycles for Walsh Modulation Generator and Walsh Filter Synthesizer respectively. In total, latency from receiving a start signal to output is 4.5 clock cycles.



\section{Walsh SID}\label{Ap:SID}
The experimental protocol referenced in~\cite{Cooper14} is summarised as follows. At $t = 0$ the qubit sensor is prepared in its initial state $\ket{\psi(0)} \equiv \ket{0}$. A resonant $\pi/2$ pulse is applied about the $\sigx$ axis, rotating the Bloch vector to the equatorial plane, coinciding an equal superposition of eigenstates, $\frac{1}{\sqrt{2}}(\ket{0}+\ket{1})$. During the subsequent time window $t\in[0,T]$ the qubit evolution is governed by a Hamiltonian term $\gamma b(t)\sigz$ describing an interaction with a dephasing field $b(t)$ (e.g. fluctuating magnetic field), with $\gamma$ an experimentally-determined coupling strength. During this time window the noisy evolution is modulated by a control field implementing a sequence of $\pi$-pulses about $\sigx$. These control-modulation pulses are synchronized with the switching times of a target Walsh function $\textsf{W}_k(t/T)$, where inclusion of the argument of the Walsh function here indicates continuous time. At the end of the acquisition time, $t = T$, a final $\pi/2$ rotation is implemented about the $-\sigy$ axis, resulting in the final state $\ket{\psi(T)}$. 

Evolution due strictly to the Walsh-modulated control (i.e. ignoring the interaction with the dephasing field) results in the final state $\ket{\psi(T)} = \ket{\psi(0)} = \ket{0}$. In this case final state readout (i.e. projective measurement along the initial state direction) yields unit fidelity, $P_k = \abs{\braket{\psi(T)|\psi(0)}}^2=1$. The dephasing field interaction introduces dephasing errors due to the accumulation of a dynamical phase $\phi_k(T)$. The reduction in fidelity due to these errors results in the general condition $P_{k} =\abs{ \braket{\psi(T)|\psi(0)}}^2 \in[0,1]$ where, for sufficiently weak noise, the characteristic value taken by the observable $P_k$ is related to the dynamical phase by
\begin{align}\label{Capellaro Probabilitiy}
P_{k} &= 
\frac{1+\sin[\phi_k(T)]}{2}. 
\end{align}
Under the digital modulation imparted by $\textsf{W}_k(t/T)$, the phase takes the form 
\begin{align}\label{Capellaro Phase}
\phi_{k}(T) = 
\gamma \int_{0}^{T}\textsf{W}_k(t/T)b(t)dt = T \gamma X_k(T)
\end{align}
where we define 
\begin{align}\label{Walsh coefficient}
X_k(T) \equiv \frac{1}{T}\int_{0}^{T}\textsf{W}_k(t/T)b(t)dt \equiv  \langle \textsf{W}_k(t/T),b(t) \rangle,
\end{align}
expressing the coefficient of $\textsf{W}_k(t/T)$ in a Walsh decomposition of $b(t)$ over the interval $t\in[0,T]$. Thus, from Eqs. \ref{Capellaro Probabilitiy}, \ref{Capellaro Phase} \& \ref{Walsh coefficient}, the Walsh coefficients are derived from the measured probability according to 
\begin{align}\label{Computed Walsh Weight}
X_n(T)  = 
\frac{\sin^{-1}[2P_n-1]}{\gamma T}.
\end{align}
Note that for $k=0$ there is no control modulation, since $\textsf{W}_0(t/T)= 1$ does not switch its value. In this case we recover the well-known Ramsey experiment, for which the accumulated quantum phase is $\phi_0(T) = \gamma \int_{0}^{T}b(t)dt$, for Ramsey time $T$. 

The resulting Walsh spectrum $\boldsymbol{X}=(X_0, X_1,...,X_{N-1})$, obtained over the single experimental window $[0,T]$, may then be used to produce a time-resolved reconstruction 
\begin{align}\label{MainEq:rec_walsh_cts1}
\hat b(t) &= 
\sum_{k=0}^{N-1}  X_k \textsf{W}_k(t/T)
\end{align}
of the stochastic signal $b(t)$. Considering the discretization of time due to the use of a fixed maximum Walsh order (corresponding to a maximum number of parallel qubit sensors), the reconstruction in Eq. \ref{MainEq:rec_walsh_cts1} is more correctly expressed in the digital representation as 
\begin{align}\label{MainEq:rec_walsh_digital}
\hat b &= 
\sum_{k=0}^{N-1}  [X_k \times \textsf{W}^{(2)}_k]
\end{align}
where the Walsh sequence $\textsf{W}^{(2)}_k$ is defined in Eq. \ref{MainEq:DigitalWalshSequence_index_k}, and $\hat b$ (argument omitted) denotes the reconstructed data stream, with consecutively streamed bits occuring once per clock cycle.

Using this prescription, we compute the fidelities $P_k$ by numerically solving the qubit time evolution. The control protocol employed in these simulations is conveniently represented in the form of a control matrix listing all relevant control operations in sequence. As an example, returning to the digital representation used in the main text, consider the Walsh sequence ${\textsf{W}_3} = (11000011)$ expressed as a binary-valued 8-bit data stream executed over a time window $[0,T]$. The minimum segment duration (time interval between consecutively streamed bits) is therefore $\tau = T/8$, with two bit-flip events occurring at $t = 2\tau, 6\tau$. The corresponding control matrix takes the form
\begin{align}\label{control matrix}
\Gamma_{3} = 
 \begin{bmatrix}
  \frac{\pi}{\tau_\pi} & \frac{\tau_\pi}{2} & \sigx \\
  0 & \tau - \frac{\tau_\pi}{2} & \hat{\mathbb{I}} \\
  0 & \tau & \hat{\mathbb{I}}\\
 \blu \frac{\pi}{\tau_\pi} & \blu \tau_\pi & \blu \sigx \\
  0 & \tau - \tau_\pi &  \hat{\mathbb{I}}\\
  0 & \tau & \hat{\mathbb{I}}\\
  0 & \tau & \hat{\mathbb{I}}\\
  0 & \tau & \hat{\mathbb{I}}\\
 \blu \frac{\pi}{\tau_\pi} &\blu \tau_\pi &\blu \sigx \\
  0 & \tau - \tau_\pi & \hat{\mathbb{I}}\\
  0 & \tau & \hat{\mathbb{I}}\\
  \frac{\pi}{\tau_\pi} & \frac{\tau_\pi}{2} & -\sigy \\
 \end{bmatrix}
\end{align}
Each row specifies the control variables associated with a unitary rotation of the Bloch vector. The first column specifies the Rabi rate of each sequential unitary, with the value 0 corresponding to a zero rotation (no control applied). The second column specifies the durations of each rotation. The third column specifies the Pauli operator associated with the axis or rotation in each unitary, with the identity $\hat{\mathbb{I}}$ corresponding to no control applied. The variable $\tau_{\pi}$ denotes the $\pi$-time of the system (time required to rotate the Bloch vector through an angle $\pi$). In the limit $\tau_{\pi}\rightarrow0$ the protocol reduces to the regime of instantaneous qubit operations assumed in Ref. ~\cite{Cooper14}. The $\pi$ pulses (shown in blue) occur at times $t = 2\tau$ and $t = 6\tau$, corresponding to the bit-flip events in $\textsf{W}_3$.

This Walsh-modulated control protocol is enacted in the presence of the fluctuating dephasing field, corresponding to the stochastic Hamiltonian term $\gamma b(t)\sigz$. The total Hamiltonian associated with this system is then substituted into the Schrodinger equation which is numerically integrated using a PDE solver package. The resulting qubit state evolution is thereby derived, enabling the computation of state fidelities $P_{k}$ at the end of the control protocol. These fidelity values are then treated as inputs for the Walsh SID module described in the main text to demonstrate its functionality. 

The Weights Estimation module employed in the Walsh SID system computes the values $X_k(T)$ by implementing Eq. \ref{MainEq:Computed Walsh Weight}. The measured probabilities, derived from simulation as above, are first converted to digital using a 13-bit sub-register of the 14-bit DAC. Each value $P_k$ is therefore resolution-limited to an integer multiple (from $0$ to $2^{13}-1$) of the minimum increment $\Delta P = 1/(2^{13}-1)\approx 1.2 \times 10^{-8}$. The 13-bit probability values are doubled using a shift register, which shifts the binary string to the left, adding a `0' to the least significant bit.
Implementing this on the FPGA does not cost any additional logic or create any delay. Input parameters need to be sampled by registers in the FPGA with the use of the system clock to synchronize the whole structure. This sampling register doubles as a shift register in the actual implementation. The shift operation, however, requires at least an $\alpha$-bit register to transform an $(\alpha-1)$-bit number (hence the choice to represent the $P_k$ in 13-bits, given a 14-bit DAC). Following the bit shift, we subtract 1 and pass the resulting 14-bit number as the input for an arcsine function on the FPGA, taking the form of a $2^{14}$-element LUT embedded in the Verilog code. This operation generally consumes vast memory on the FPGA, however this method provides the highest speed solution for the arcsine function, since it consumes only one clock cycle to complete the operation. The result is a value for the numerator in Eq. \ref{MainEq:Computed Walsh Weight}. The output of the arcsine function is fed into the divider to adjust the amplitude due to the change of coupling strength coefficient ($\gamma$) and acquisition time frame (T).  The divider will adjust the amplitude of dividend $\gamma TX_k$ with the divisor $\gamma T$ and can handle up to 14 bits of both dividend and divisor. The quotient is in form of 14-bit output integer. This output integer is thereafter the input data for the DAC.

\end{document}